# Design and Performance of an InGaAs–InP Single-Photon Avalanche Diode Detector

Sara Pellegrini, Ryan E. Warburton, Lionel J. J. Tan, Jo Shien Ng, Andrey B. Krysa, Kristian Groom, John P. R. David, Sergio Cova, Michael J. Robertson, and Gerald S. Buller

*Abstract*—**This paper describes the design, fabrication, and performance of planar-geometry InGaAs–InP devices which were specifically developed for single-photon detection at a wavelength of 1550 nm. General performance issues such as dark count rate, single-photon detection efficiency, afterpulsing, and jitter are described.**

## I. Introduction

SINGLE-PHOTON counting and single-photon timing have become increasingly important in a number of applications such as time-resolved photoluminescence [1], optical time-domain reflectometry (OTDR) [2] and time-of-flight laser ranging [3] and imaging. More recently they have been employed in quantum key distribution [4] and noninvasive testing of VLSI circuits [5]. Commercially available InGaAs-InP avalanche photodiodes (APDs) designed for use in linear multiplication mode have been tested in Geiger mode [6], [7] in order to extend the spectral range of efficient single-photon detection to wavelengths greater than that afforded by the currently available Si-based single-photon avalanche diode (SPAD) detectors (i.e., wavelengths greater than 1 μm). These InGaAs–InP detectors have exhibited good single-photon detection efficiency (SPDE) (SPDE > 10%) and subnanosecond timing jitter, however they have limited counting rates due to the severe deleterious effects of the afterpulsing phenomenon. Alternative approaches to photon-counting have also been developed such as quantum-dot (QD) field effect transistors [8] and QD resonant tunnelling diodes [9] which have shown promising results in terms of SPDE at wavelengths around 684 nm [10]. Superconducting single-photon detectors [11] have demonstrated a SPDE of approximately 5% at a wavelength of 1.55 μm when cooled to a temperature of 4 K.

This paper describes the design, fabrication and characterization of planar geometry InGaAs–InP avalanche diode detectors, which have been specifically designed for single-photon detection. We describe the temperature dependence of the device SPDE, dark count rate (DCR), noise equivalent power (NEP), and jitter, focusing on the effect of the critical InGaAsP grading layer between the narrow-gap InGaAs absorption layer and the wide-gap InP multiplication layer. This study represents a fabrication program for planar InGaAs–InP SPADs and highlights some important issues in device design.

## II. Device Structure

The devices studied in this paper are of planar geometry and are based on the separate absorption, grading and multiplication region APDs [12] commonly used for applications requiring internal gain of weak input optical signals. In this structure the optical absorption is performed in the $In_{0.53}Ga_{0.47}$ As, which is not suitable for multiplication, since tunnelling in this material [13] would take place at a lower electric field than that required for impact ionization. Instead, multiplication takes place in the wider gap InP. In order to smooth the large valence band discontinuity between InGaAs and InP (see below), a suitable quaternary grading layer of InGaAsP is added between the two materials.

It is important to notice that the design criteria for the devices described in this paper are similar to linear multiplication APDs except in three important respects. 1) The design must allow low temperature operation (at least down to 150 K) in order to reduce the DCR due to thermally generated carriers, therefore the difference between the punch-though voltage and the breakdown voltage must be of at least 30 V, since the former remains almost unchanged, while the latter is reduced with decreasing temperature. 2) The design must take into account the higher electric fields used in comparison with linear multiplication devices. In particular, it is necessary to avoid as far as possible tunneling effects, even those with low probability, since they will also increase the DCR. This is achieved by having a low electric field in the InGaAs. 3) The mask design incorporates different diffused area curvatures in order to accommodate the smaller volume structures needed for low DCRs whilst avoiding edge breakdown effects, such as the 10-μm diameter devices (see later). The specific choice of the zinc diffused planar geometry was based on previous studies done by our research groups, which highlighted consistently lower DCRs in devices fabricated using this approach [6], [7], [14].

Fig. 1(a) is a schematic showing the principle of device operation in relation to its energy band diagram. When a photon is absorbed in the InGaAs layer, an electron–hole pair is generated. The electron is swept toward the back contact, while the hole drifts in the depleted InGaAs layer toward the InP multiplication region where it undergoes impact ionization. The intermediate bandgap InGaAsP is introduced between InGaAs and InP to ensure efficient transfer of holes across the large valence band discontinuity. In this paper we will show the effect of the grading layer composition on the SPDE of the InGaAs–InP SPAD.

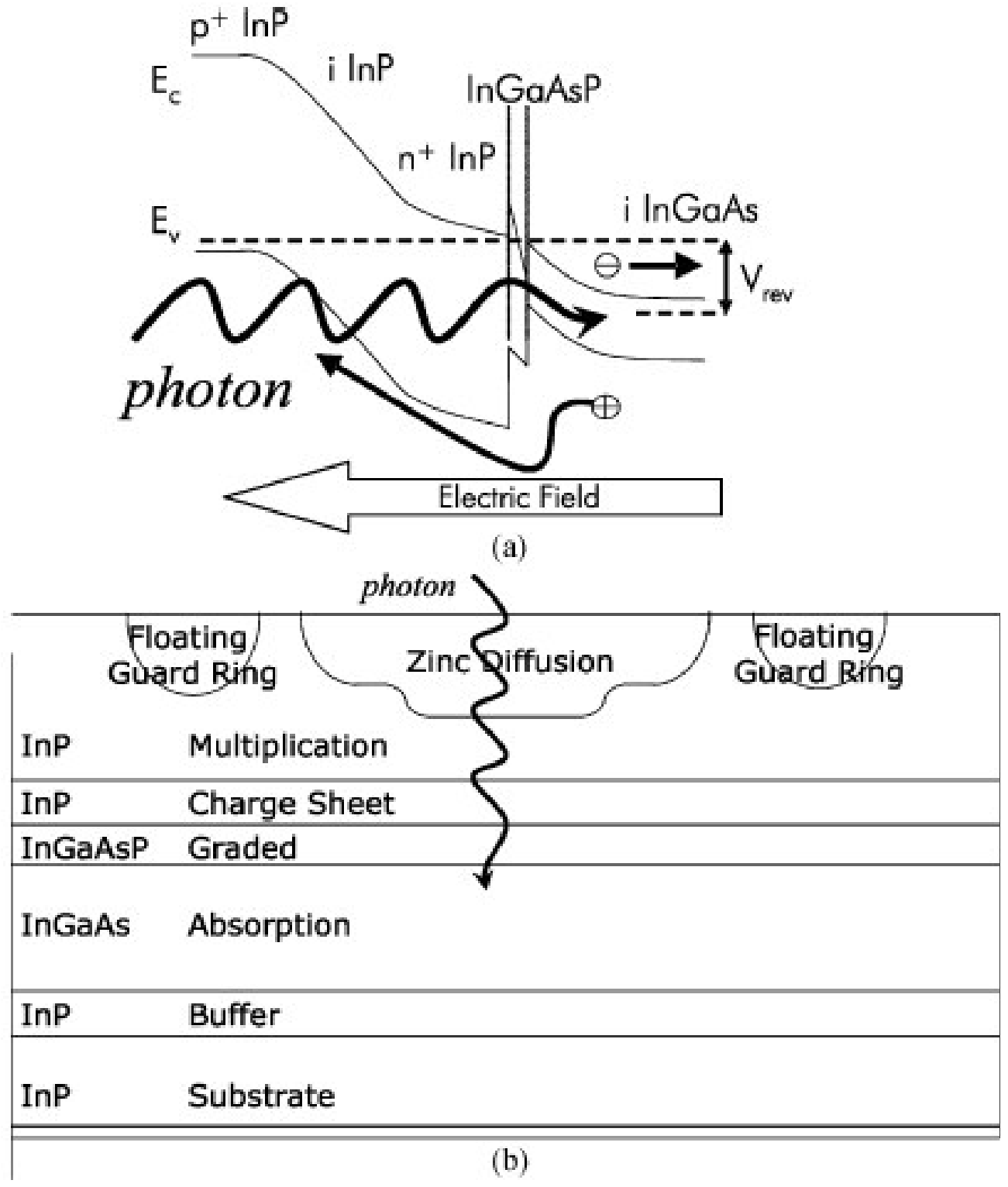

Fig. 1. (a) Energy band diagram of the device under reverse bias conditions. (b) Schematic cross section of a planar SPAD with one floating guard ring, a top p-contact to the active area, and a bottom n-contact to the substrate.

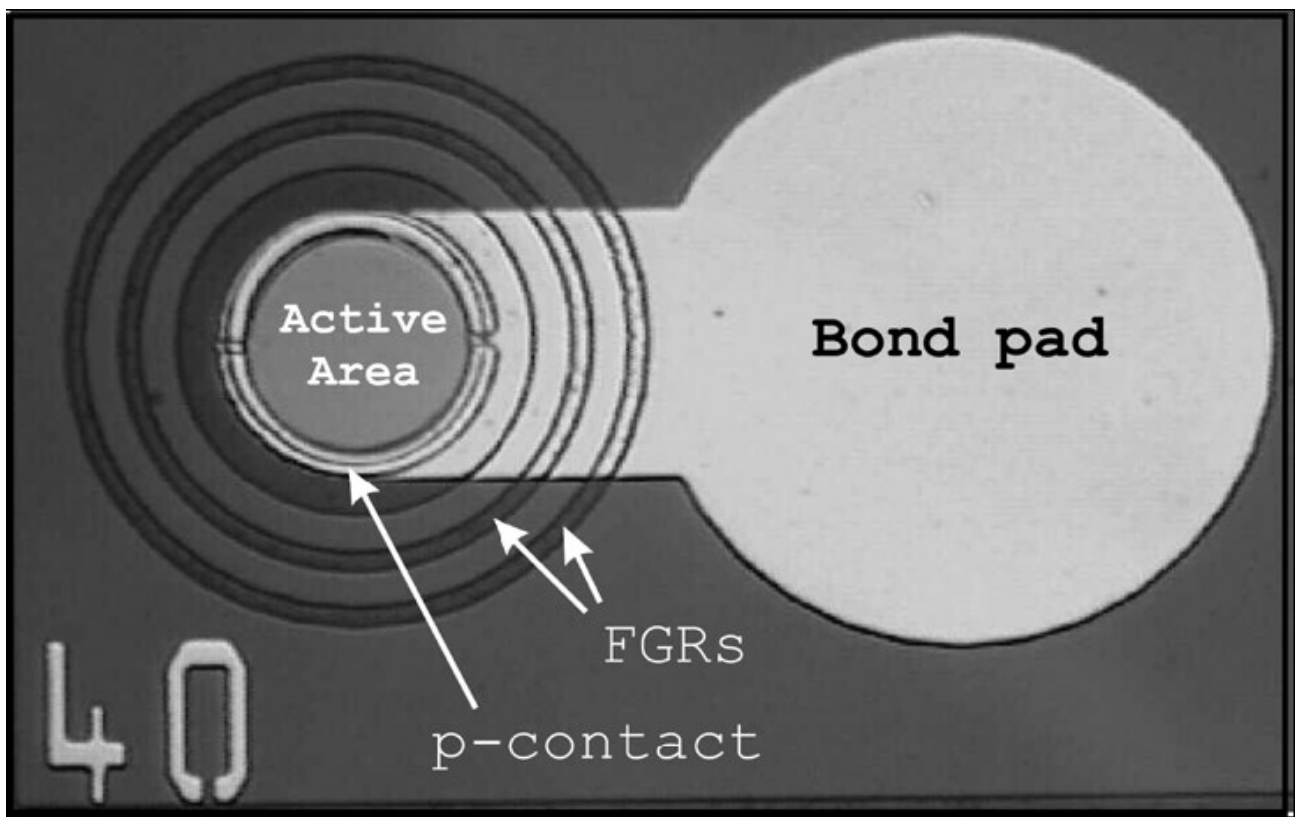

Fig. 2. Microscope image of a 40-_m-diameter InGaAs–InP SPAD.

The device material was grown epitaxially by metal organic chemical vapor deposition. As shown in Fig. 1(b), the device consists of a 2.5-μm-thick layer of lightly n-doped InGaAs for efficient absorption of photons at a wavelength of 1.55 μm, and a 1- m-thick layer of nominally undoped InP for multiplication to take place. The charge sheet layer is a 300-nm-thick InP n-doped at $6 \times 10^{16}$ $cm^{-3}$, which is needed to reduce the electric field from the InP multiplication region to the InGaAs absorption layer. For the grading layer, two main device designs were grown: 1) with a graded region consisting of one quaternary with an exactly intermediate bandgap between InGaAs and InP, which we shall call SPAD-1Q and 2) with a graded region composed of three sublayers of stepped bandgap, which we shall call SPAD-3Q. Our previous attempts to utilize a continuously varying gap were not successful. The p-n junction is fabricated by diffusing the p-type dopant zinc into the top InP region. In order to minimize the DCR of the device, a planar geometry device was formed using the diffusion of the p-type Zn dopant alone. The use of etch and regrowth techniques [15], [16] was avoided in order to reduce interface states that may contribute to dark count and afterpulse levels. The active area is shaped by the diffusion of zinc via two separate processes using different mask sets as first suggested by Liu *et al.* [12]. This double diffusion reduces the curvature at the edge of the device, thus reducing the likelihood of edge breakdown [17], where the device will preferentially exhibit avalanche breakdown at the edge of the active area rather than in the central region, hence significantly lowering the SPDE. The likelihood of edge breakdown can be further reduced by the introduction of floating guard rings (FGR) [12] that decrease the electric field at the edges of the main junction.

An optical microscope image of the device prior to bonding is shown in Fig. 2. The active area of the device is defined by the opening for the deeper zinc diffusion, and the metal contact. Both the central diffusions and the floating guard rings are concentric. The contact is formed by evaporated Au-Zn-Au for the top p-side, and InGe–Au for the n-substrate.

## III. Device Characteristics

The reverse current–voltage characteristics showed a similar behavior for both structures. The only evident difference in the curves shown in Fig. 3 for the dark current measured in 10- m-diameter devices from both structures at room temperature is the position of the punch-through voltage ($V_{PT}$). This is the voltage at which the depletion region extends into the In-GaAs layer and efficient collection of the carriers generated by absorption of 1.55-μm wavelength light can take place. The dark current ($I_D$) and breakdown voltages ($V_{BD}$) are the same for both structures. The dark current at room temperature was measured to be 3 nA at 95% of $V_{BD}$, and $V_{BD}$ varied between 105 and 110 V across approximately 10 mm of wafer. The difference in the punch-through voltage value is most likely due to the slight thickness variation between the two structures in the top InP layer. From Fig. 4, it can also be noticed that the punch-through voltage does not change significantly with temperature, as expected, nor does the value of the photocurrent measured just above punch-through with an incident light of 1.55-μm wavelength. The breakdown voltage decreases linearly with temperature with a coefficient of ~0.17 V/K, which is consistent with previously published results [7].

For the characterization of these detectors in terms of photon-counting performance, we used the setup shown in

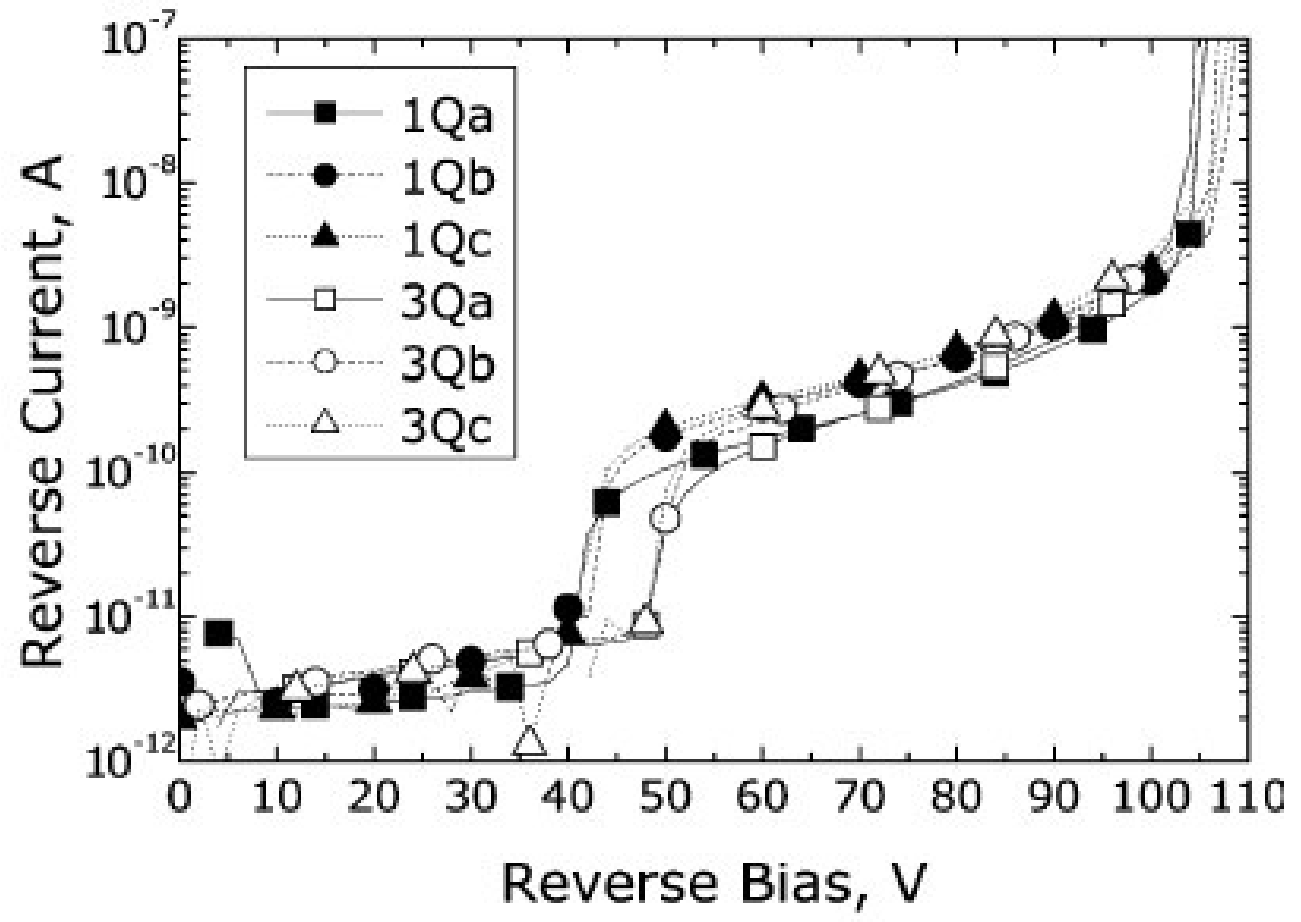

Fig. 3. Reverse dark current measured in a selection of 10-_m-diameter devices from both structures SPAD-1Q and SPAD-3Q at room temperature.

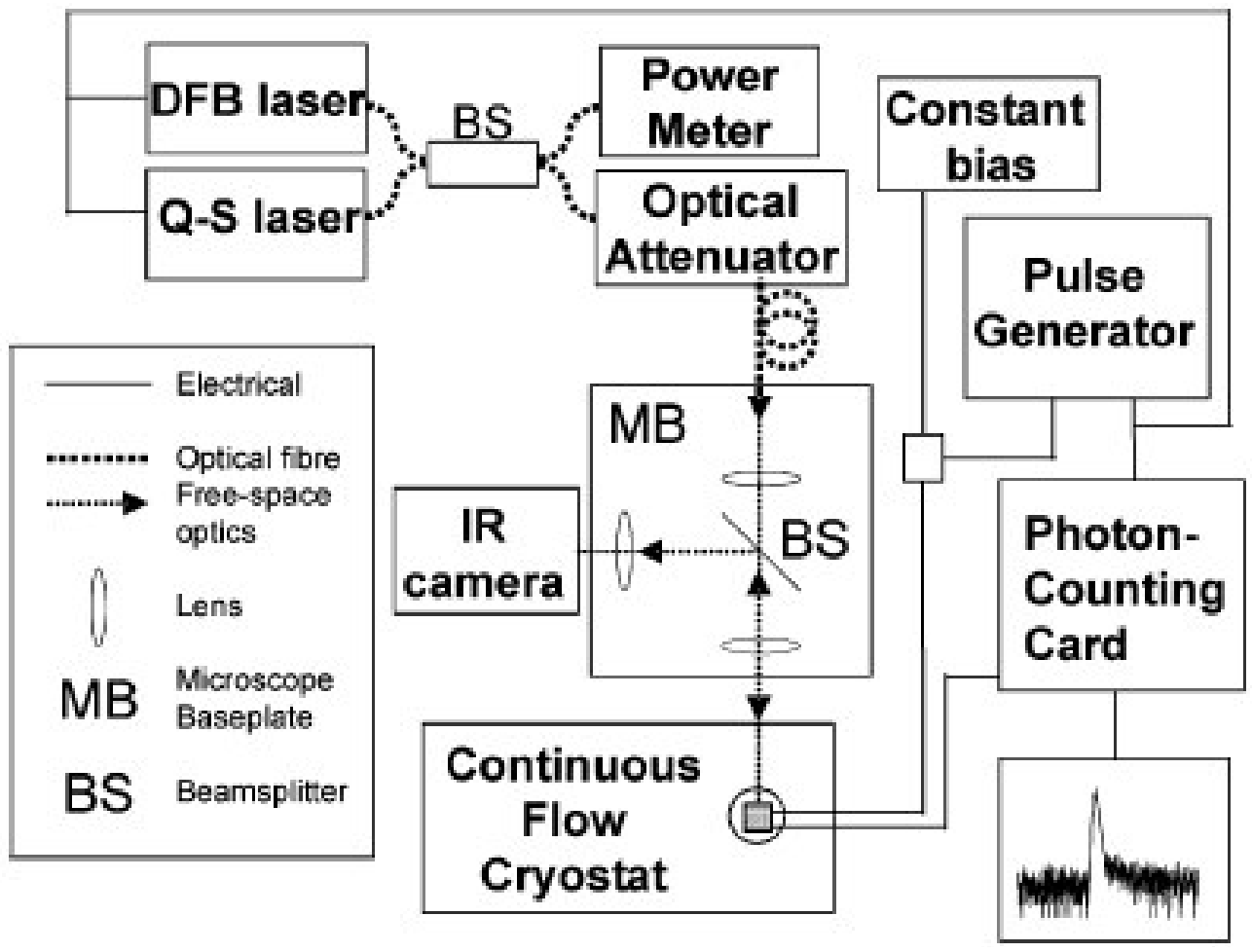

Fig. 5. Experimental setup for photon-counting characterization of the InGaAs–InP SPAD. MB corresponds to the microscope baseplate, and BS is a pellicle beamsplitter.

Fig. 5. Two lasers were used for the characterization: a passively *Q*-switched diode laser [18] for the timing measurements with a low jitter <20 ps pulsewidth and a distributed feedback (DFB) laser for alignment and SPDE measurements. This setup allowed for SPDE and jitter measurements to be performed without altering the optical alignment of the system. One output of the beamsplitter was used to constantly monitor the laser power, the second was heavily attenuated to ensure an average of less than one photon per pulse arrived at the SPAD detector. The output fiber of the optical attenuator was attached to a microscope baseplate [19] where it was collimated and then focused on the sample using a ×20 magnification microscope objective lens. The microscope baseplate housed all optical components for imaging and delivering the laser to the detector, allowing high-precision alignment and focusing on the detector. White light and an infrared camera were used to image the devices. The sample was fixed in a continuous flow cryostat under high vacuum, and was connected to two coaxial cables for biasing. The dc bias had a gate superimposed upon it by the pulse generator to take the device beyond avalanche breakdown and into the Geiger regime, i.e., the gated mode operation [20]. This pulse generator also provided the synchronization pulse for the laser drivers and signals the start for the timing of the photon-counting card. The avalanche pulse current was small and therefore amplified before passing through a constant fraction discriminator. The photon-counting statistics were measured using an Edinburgh Instruments Ltd. TCC900 photon-counting PC card.

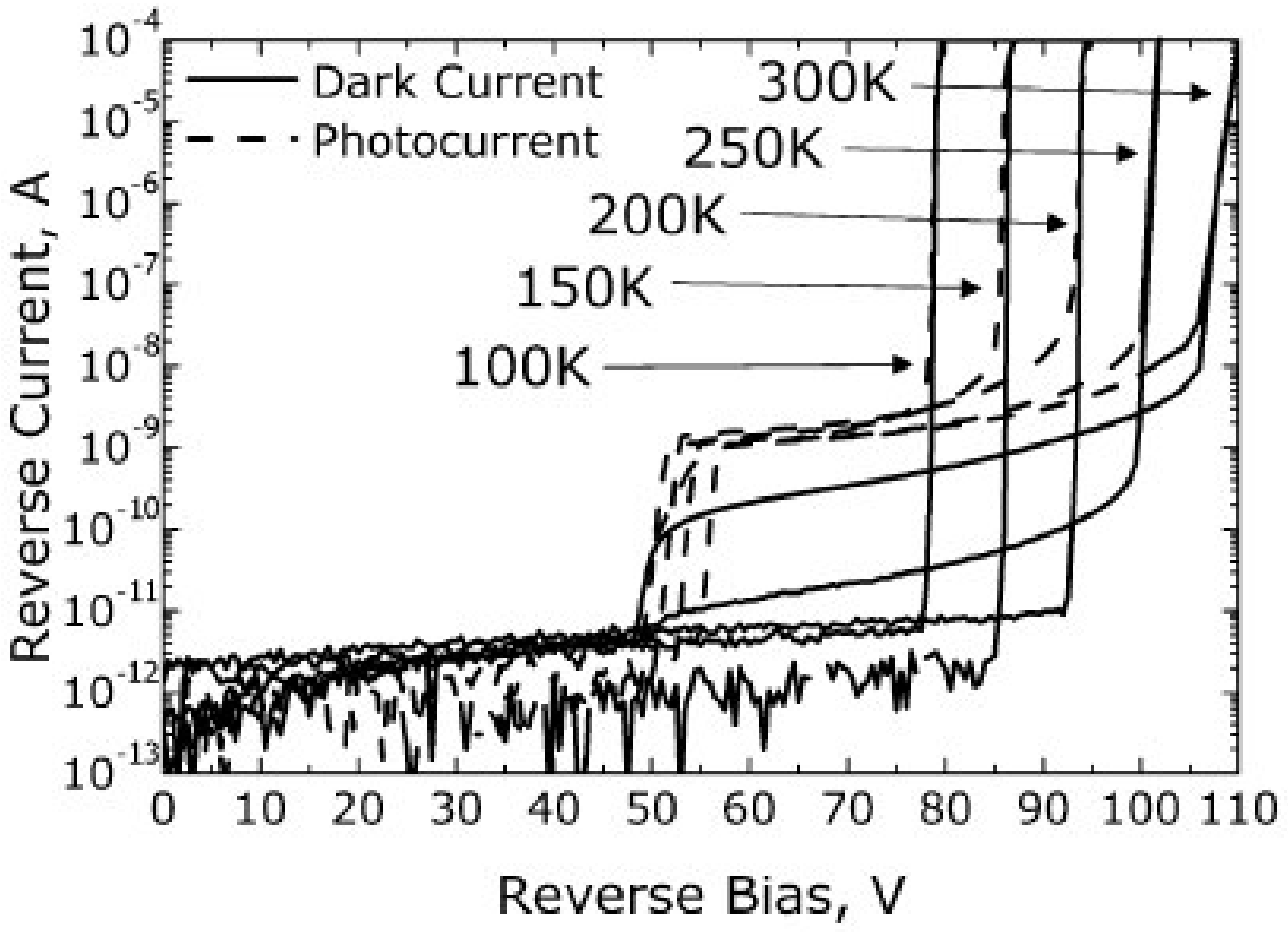

Fig. 4. Dark current (solid lines) and photocurrent (dashed lines) measurements at different temperatures for a 10—_m-diameter device from the SPAD-3O structure.

Here, we report the device characterization in terms of the figures of merit that are relevant for single-photon detection such as DCR, SPDE, NEP and timing jitter. The DCR is the count rate of the detector when no light impinges on it, it is due to thermally generated carriers and is enhanced by the release of trapped carriers. The SPDE is the probability that a photon incident on the detector generates an output pulse. It can be measured as the ratio of the number of detected photon events compared to the number of incident photons averaged over a large number of optical pulses, and assuming a low probability of multiple photons in each optical pulse. The NEP is a figure of merit for photodetectors that takes into account both the SPDE and the DCR. It represents the least measurable optical power (with 1-Hz bandpass noise filtering, that is, 1-s total counting time in single-photon counting) and is given by the following equation:

$$NEP = \frac{h\nu}{SPDE}\sqrt{2DCR} \qquad (1)$$

where $h\nu$ is the single-photon energy. The timing jitter is the full-width at half-maximum of the SPAD response and is measured in a time-correlated photon counting (TCPC) set up when the detector is illuminated with a highly attenuated pulsed laser with shorter pulse duration, in practice approximately 20 ps. All the results taken on the InGaAs–InP SPADs at different temperatures are presented in comparison with the best commercially available linear multiplication APDs (Epitaxx EPM239AA and Fujitsu FPD15W5) operated in Geiger mode. The detectors were operated in gated mode which, in the measurements described in this paper, consisted of applying a dc bias to the detector at 1 V below $V_{BD}$, and applying a gate of 100-ns duration and an amplitude corresponding to applying an excess bias (i.e., bias above $V_{BD}$) of 10% at all the measured temperatures. The value of 10% was chosen for all the devices, this being the value at which the SPDE is high, although still below the saturation level, while the DCR is low enough to guarantee a good NEP. The time period between the gates was 100 µs, corresponding to a repetition rate of 10 kHz for all the measurements, to avoid the effects of afterpulsing. The Epitaxx EPM239AA has a nominal active area diameter of 40 m and some information regarding its microstructure has been published previously [21], [22]. This published material indicates that the p-well is obtained by two different zinc diffusion processes. The edge breakdown is prevented by use of two floating guard rings at each side of the well. The grading of the quaternary layer is obtained by growing three different sublayers of varying composition of InGaAsP. Substrate entry geometry allows incident photons a second chance at absorption after reflection at the annealed AuZn top contact, thus permitting a thinner InGaAs absorption layer thickness and a corresponding reduction in the thermally generated leakage current for a given device diameter. This device has a room temperature dark current at 95% $V_{BD}$ of approximately 100 pA.

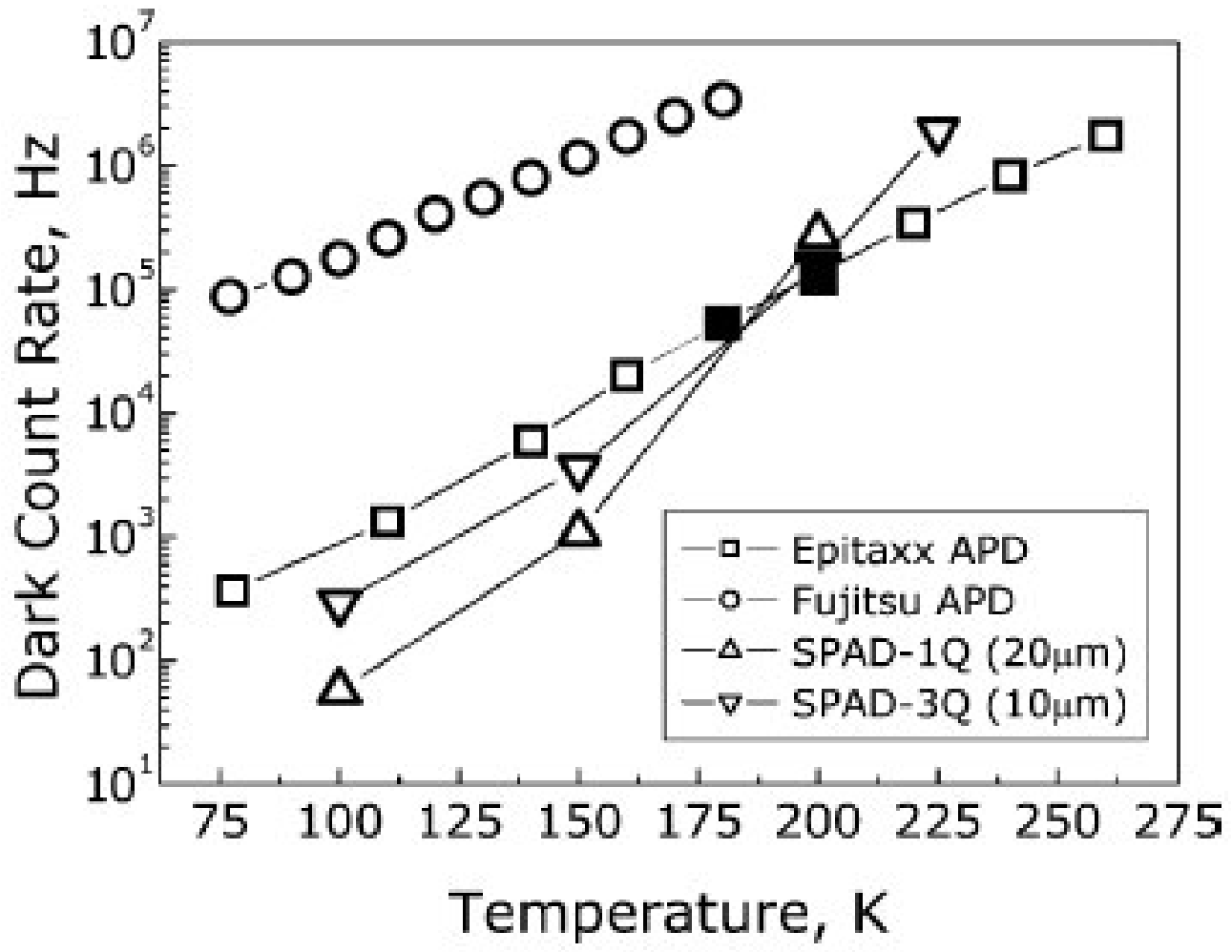

Fig. 6. Comparison of the DCR as a function of temperature of the Fujitsu APD, the Epitaxx APD, and both the SPAD-1Q and SPAD-3Q.

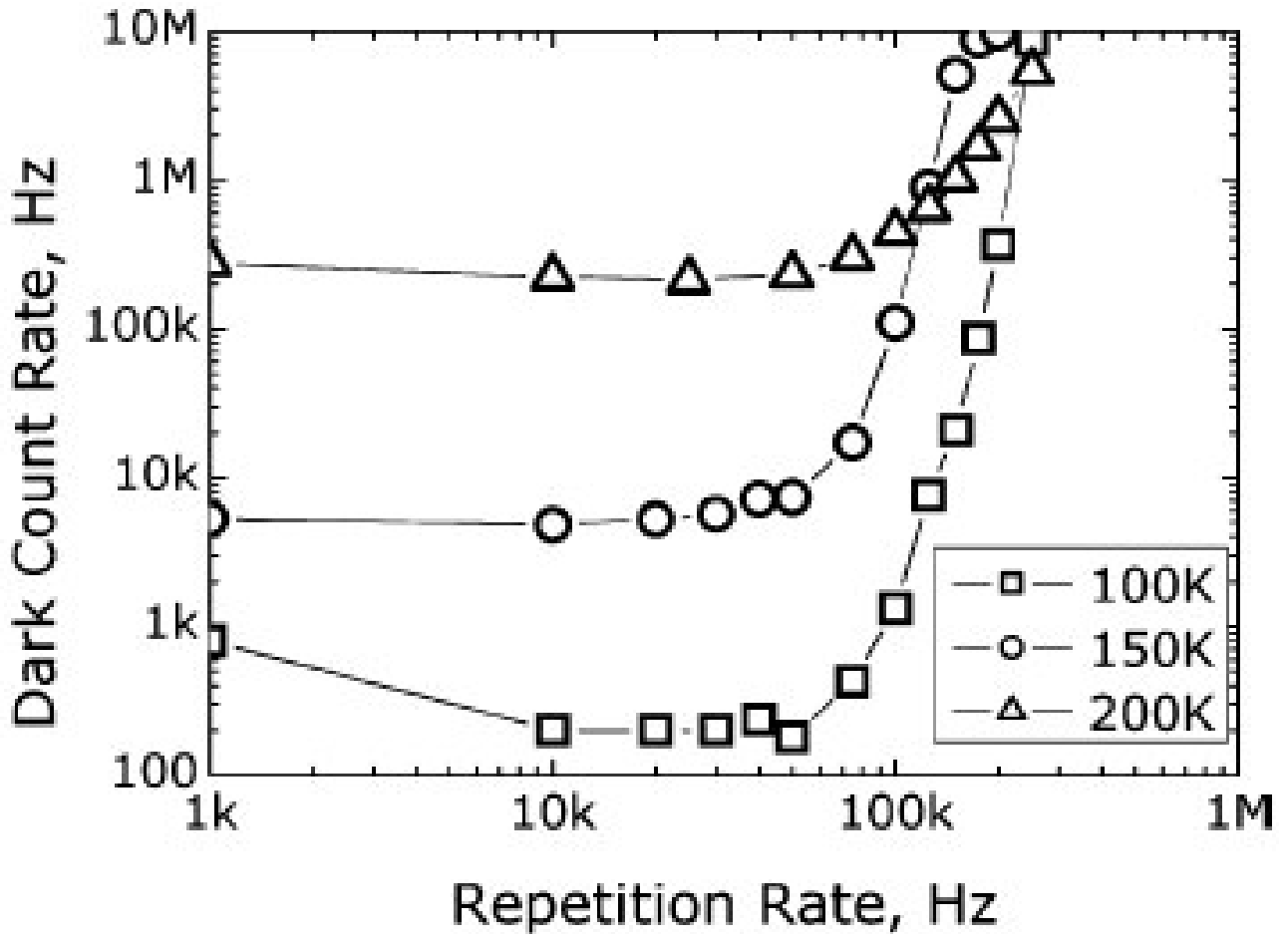

Fig. 7. DCR as a function of the repetition rate for a 20-µm-diameter SPAD-3Q at different temperatures when a 100-ns duration gate is applied. The DCR increases rapidly with repetition rate for repetition rates higher than 50 kHz.

The Fujitsu FPD15W5 is a free space device with a diameter of 80 µm. According to the literature [15], [16], this APD consists of a separate absorption, grading and multiplication structure, again with three sublayers used for the InGaAsP grading layer. The main difference with respect to the Epitaxx APD is the geometry employed for suppression of edge breakdown. The Fujitsu devices' active region is formed by one single diffusion and the edge breakdown is prevented by incorporating two lateral charge sheets. This can be achieved by etching and regrowth of the device, which appears to be the key fabrication difference between the two commercially sourced devices. This processing step, however, has an adverse influence on the quality of the material in the active junction and, in fact, the dark current at room temperature has a much higher value than the Epitaxx device, being in the order of 1 nA.

Fig. 6 shows the DCR of the detectors as a function of temperature, compared with selected linear multiplication devices operated in Geiger mode. The measurements are taken at the same bias levels used for comparison of the SPDE shown in Fig. 8. Generally, the SPAD devices presented here compare well in terms of DCR with the selected APD devices. The expected behavior of the DCR as a function of temperature is an exponential dependence. At temperatures higher than 175K the SPADs have a higher DCR than the Epitaxx APD, but their DCR drops at lower temperature, and more so for the SPAD-1Q. This is most likely due to thermally generated holes in the InGaAs not being efficiently swept across the valence band discontinuity at low temperature due to their reduced thermal energy. This effect is more significant in the SPAD-1Q due to the higher valence band discontinuity between adjacent sublayers in comparison to the SPAD-3Q device.

Fig. 7 shows the DCR dependency on repetition rate to illustrate the effect of the afterpulsing phenomenon. It is evident that at lower repetition frequencies (i.e., <50 kHz) with a 100-ns duration gate, the DCR remains constant. This indicates that the main contribution to the DCR is given by thermally generated initiating carriers, rather than initiated by carriers trapped during previous avalanche events. At higher repetition rates, carriers trapped in deep levels during an avalanche are subsequently released during the following gate-on period, triggering a new avalanche. This generates an increase in the DCR known as the afterpulsing effect. In order to operate at higher repetition rates, a shorter gate could be applied [23] to reduce the total charge in the avalanche pulse. For comparing the true primary DCR, we will avoid the enhancement due to afterpulsing and consider only thermally generated carriers by using a low repetition rate of 10 kHz.

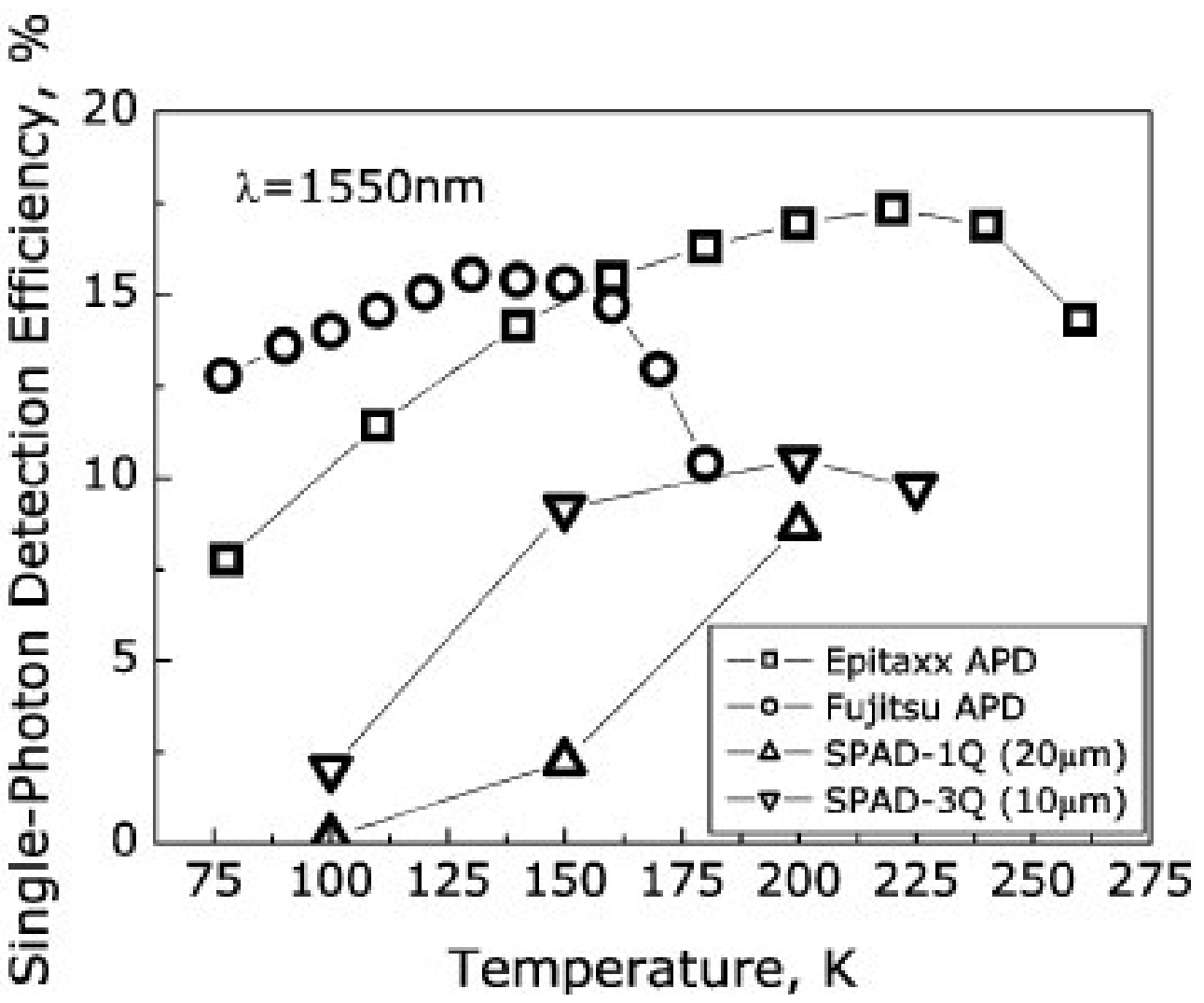

Fig. 8. Comparison of SPDE at different temperatures for the Epitaxx APD, the Fujitsu APD, and the SPAD-1Q and SPAD-3Q.

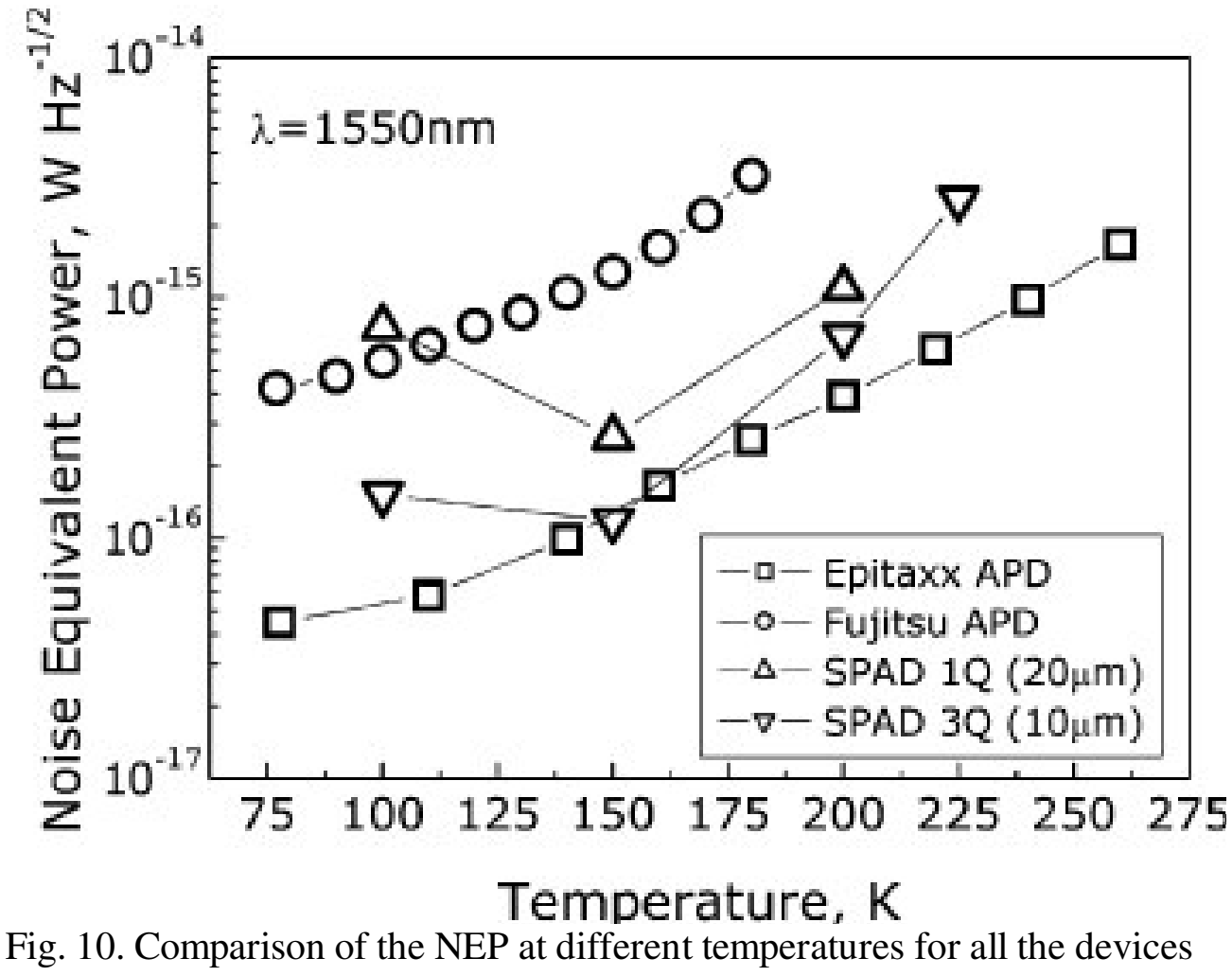

Fig. 10. Comparison of the NEP at different temperatures for all the devices described in this paper.

The device SPDE measurements as a function of temperature with a 10% relative excess bias applied are shown in Fig. 8. At high temperatures, the SPADs reach a SPDE of about 10%, which is comparable to the best reported performance of InGaAs–InP single-photon detectors operating in comparable conditions. Higher SPDE was reported only for operation withultrashort gate duration (about 1 ns) and larger excess bias [23]. When the temperature is reduced, the detection efficiency drops significantly in both the SPAD-1Q and SPAD-3Q detectors, but significantly more so for the SPAD-1Q detectors. Such a temperature dependence is much less evident with the commercially available detectors shown in the diagram for comparison.

By plotting the SPDE in an Arrhenius plot, as shown in Fig. 9 we measured an activation energy for the thermionic emission of the holes of 70 meV for the SPAD-1Q and 40 meV for the SPAD-3Q. These values are consistent with those calculated by Forrest *et al.* [24] for the valence band discontinuity when a reverse bias of approximately 90 V is applied to the detector. If we assume evenly spaced bandgaps in the SPAD-3Q structure, we expect a maximum valence band discontinuity approximately half that of the SPAD-1Q structure. This was confirmed by experimental results as shown in Fig. 9. Despite the clear improvement achieved by adding two more bandgap steps to the grading layer, further development of the quaternary layer will be needed to attain the value of 8.8 meV measured for the Epitaxx APD. Also to further improve the SPDE a higher electric field should be used at the hetero-interfaces to improve the transport efficiency of the photogenerated holes. Moreover, an antireflection coating together with a back entrance geometry similar to the Epitaxx devices could be employed for improving the detector performance.

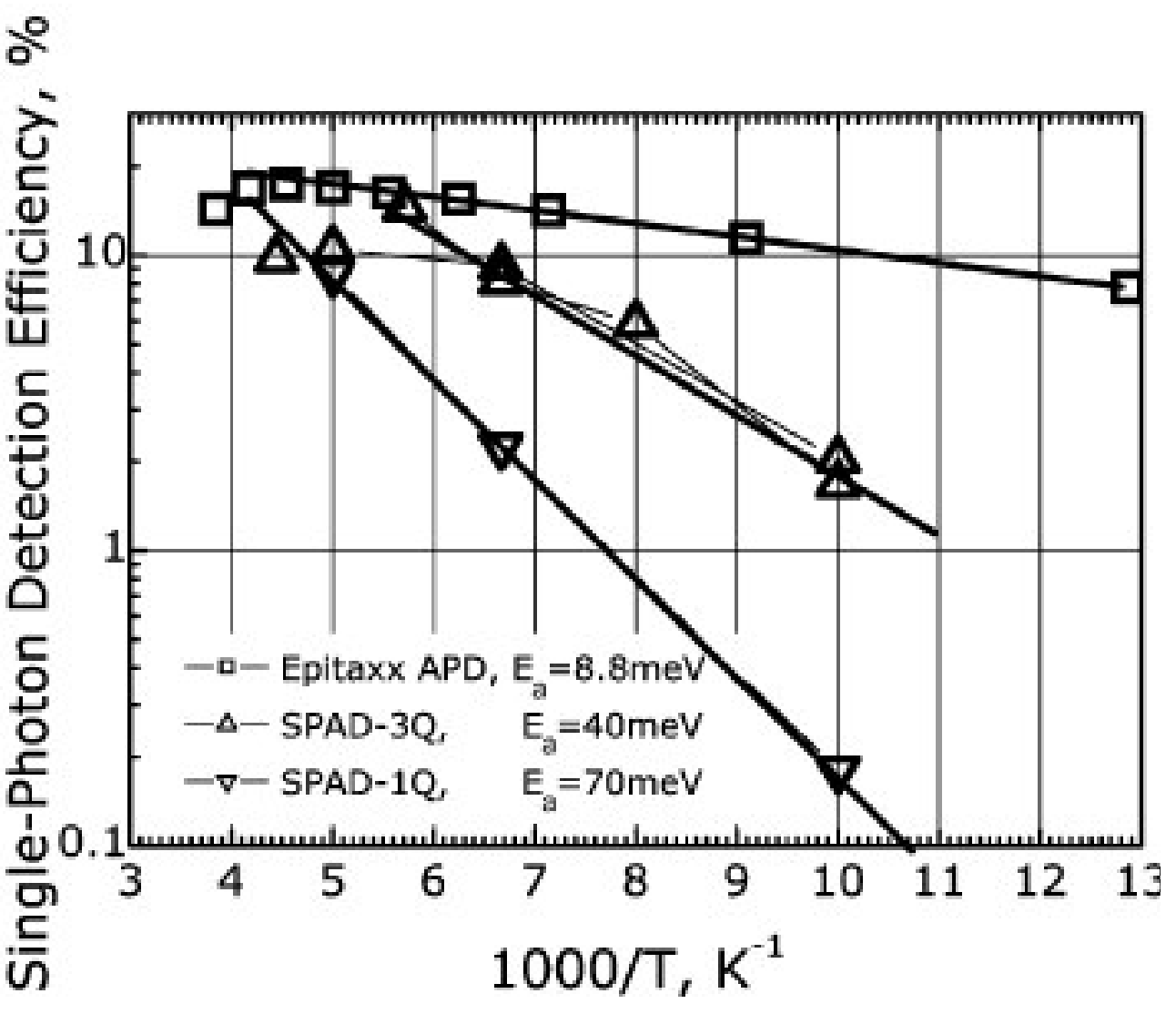

Fig. 9. Arrhenius plot of the SPDE for the Epitaxx APD and the two different SPAD structures.

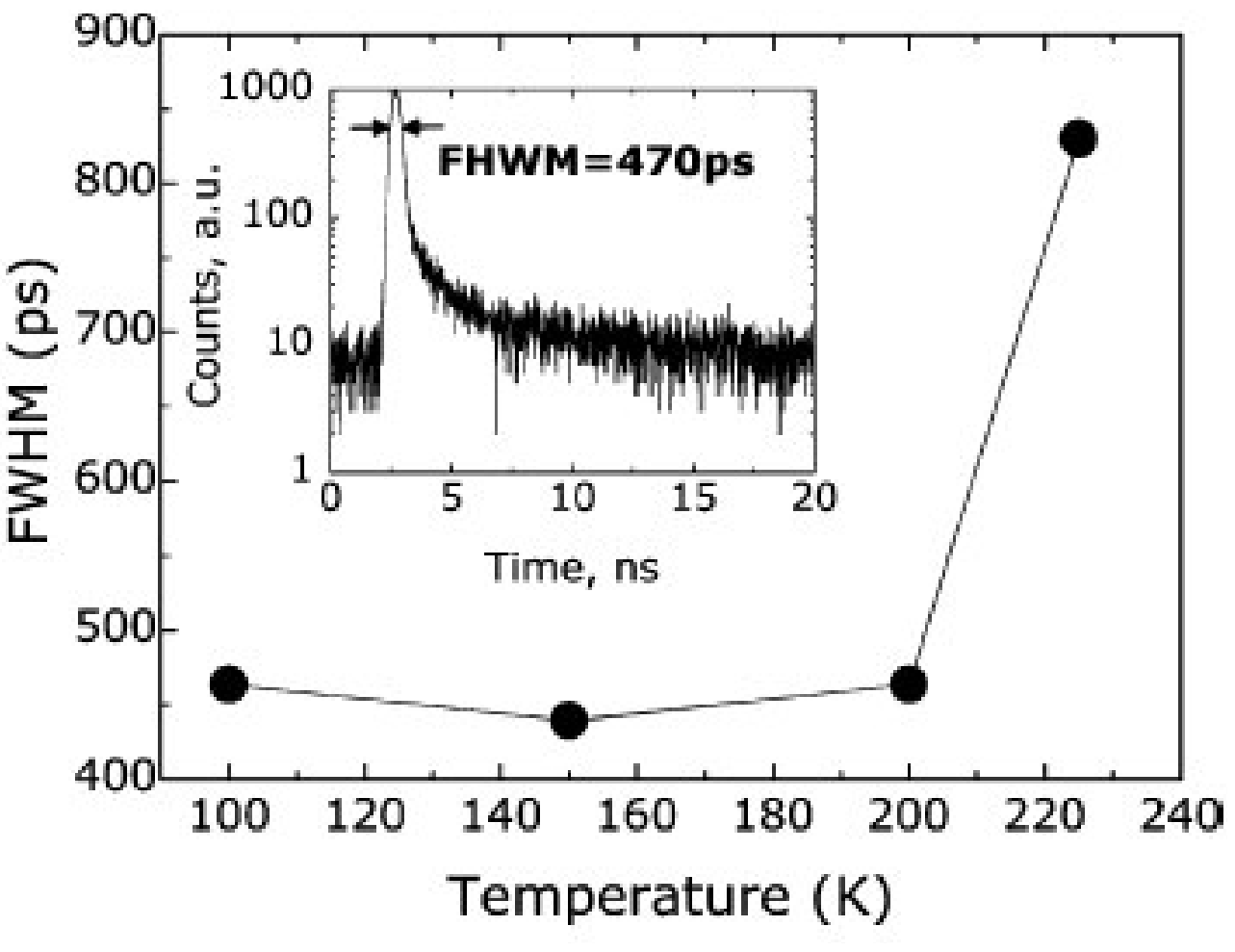

Fig. 11. Timing jitter of a 10-_m-diameter SPAD-3Q at different temperatures. The inset shows the raw data histogram for one of the measurements.

Fig. 10 shows the results in terms of NEP. For a device to be good, the NEP needs to be as low as possible, therefore, at high temperature the SPADs still compare well with the commercial devices, but their performance degrades at lower temperature, due to the strong reduction in SPDE.

As far as the timing jitter is concerned, the detectors showed a measured response of approximately 450 ps at best (this includes the measurement system response), as shown in Fig. 11. The jitter measured at 225 K increased considerably to 800 ps, since the output signal of the SPAD is small (20 mV) and the signal-to-noise ratio is poor (3:1). The low signal-to-noise of the analogue output signal could be significantly improved by improved contacts to reduce the series resistance of the device. The noise inherent in the output circuitry is also partly due to particularly long cables used within the cryostat, which make a significant contribution to the jitter measured in these experiments. The measured values did not appear to depend on temperature, and could be improved by employing more advanced electronics for the extraction of the SPAD current output and improved low-temperature packaging of the device. Nevertheless a jitter of 450 ps is reasonable in the context of quantum key distribution applications at wavelengths of 1.3 and 1.55 μm.

## IV. Conclusions

We have successfully developed SPADs for efficient detection at a wavelength of 1.55 μm with a SPDE of 10% and a NEP of $6 \times 10^{-16}$ W $Hz^{-1/2}$ at 200 K. It is clear that considerable potential exists for further improvements by 1) the use of antireflection coatings, which should improve the SPDE by approximately 30%; 2) further refinement of the quaternary layers as discussed above; 3) enhance the electric field at the hetero-interface; 4) further refinement of the Zn diffusion front to more fully centralize the active area of the device; and 5) the effect of changing the thickness of the avalanche junction may be worth investigating, but in direction opposite to that of current research on amplifying APDs. Very narrow avalanche regions are employed for reducing the multiplication noise in APDs, but in SPADs they would reduce the SPDE without bringing significant advantages. Thicker avalanche regions could instead be experimented in SPADs for obtaining higher detection efficiency with lower electric field, possibly without degrading the timing jitter and the dark counting rate. We also have investigated the effect of the grading layer on the SPDE of a custom built SPAD, and fully tested the devices for robustness in temperature cycling and long operation times. The measured timing jitter of 450 ps could be reduced through improved ohmic contacts to the detectors, as well as improved packaging.

The devices still suffered from a level of afterpulsing comparable to that of linear multiplication devices previously studied. However, these devices have reached a sufficiently high level of operating performance that they can be used as a basis for further analysis and a subsequent development program to investigate the origin and methods of removal of the trap states that cause the afterpulsing phenomenon

## References

[1] G. S. Buller, S. J. Fancey, J. S. Massa, A. C. Walker, S. Cova, and A. Lacaita., “Time-resolved photoluminescence measurements of In-GaAs–InP multiple-quantum-well structures at 1.3 μm wavelengths by use of germanium single-photon avalanche photodiodes,” *Appl. Opt.*, vol. 35, no. 6, pp. 916–921, 1996.

[2] A. L. Lacaita, P. A. Francese, and S. Cova, “Single-photon optical-time-domain reflectometer at 1.3 μm with 5 cm resolution and high sensitivity,” *Opt. Lett.*, vol. 18, no. 13, pp. 1110–1112, 1993.

[3] S. Pellegrini, G. S. Buller, J. M. Smith, A. M. Wallace, and S. Cova, “Laser-based distance measurement using picosecond resolution time-correlated single-photon counting,” *Meas. Sci. Technol.*, vol. 11, pp. 712–716, 2000.

[4] K. J. Gordon, V. Fernandez, P. D. Townsend, and G. S. Buller, “A short wavelength gigaHertz clocked fiber optic quantum key distribution system,” *IEEE J. Quantum Electron*., vol. 40, no. 10, pp. 900–908, Oct. 2004.

[5] F. Stellari, A. Tosi, F. Zappa, and S. Cova, “CMOS circuit testing via time-resolved luminescence measurements and simulations,” *IEEE Trans. Instrum. Meas*., vol. 53, no. 1, pp. 163–169, Jan. 2004.

[6] P. A. Hiskett, G. S. Buller, A. Y. Loudon, J. M. Smith, I. Gontijo, A. C. Walker, P. D. Townsend, and M. J. Robertson, “Performance and design of InGaAs–InP photodiodes for single-photon counting at 1.55 μm,” *Appl. Opt.*, vol. 39, no. 36, pp. 6818–6829, 2000.

[7] A. Lacaita, F. Zappa, S. Cova, and P. Lovati, “Single-photon detection beyond 1 μm: Performance of commercially available InGaAsInP detectors,” *Appl. Opt.*, vol. 35, pp. 2986–2996, 1996.

[8] B. E. Kardynal, A. J. Shields, N. S. Beattie, I. Farrer, K. Cooper, and D. A. Ritchie, “Low-noise photon counting with a radio-frequency quantum-dot field-effect transistor,” *Appl. Phys. Lett.*, vol. 84, no. 3, pp. 419–421, 2004.

[9] J. C. Blakesley, P. See, A. J. Shields, B. E. Kardyna, P. Atkinson, I. Farrer, and D. A. Ritchie, “Efficient single photon detection by quantum dot resonant tunneling diodes,” *Phys. Rev. Lett.*, vol. 94, no. 6, p. 067401, 2005.

[10] N. S. Beattie, B. E. Kardynal, A. J. Shields, I. Farrer, D. A. Ritchie, and M. Pepper, “Single-photon detection mechanism in a quantum dot transistor,” *Physica E*, vol. 26, no. 1–4, pp. 356–360, 2005.

[11] G. N. Gol’tsman, O. Okunev, G. Chulkova, A. Lipatov, A. Semenov, K. Smirnov, B. Voronov, A. Dzardanov, C. Williams, and R. Sobolewski, “Picosecond superconducting single-photon optical detector,” *Appl.. Phys. Lett.*, vol. 79, no. 6, pp. 705–707, 2001.

[12] Y. Liu, S. R. Forrest, M. J. Lange, G. H. Olsen, and D. E. Ackley, “A planar InP/InGaAs avalanche photodiode with floating guard ring and double diffused junction,” *J. Lightw. Technol.*, vol. 10, no. 2, pp. 182–193, Feb. 1992.

[13] S. R. Forrest, M. DiDomenico Jr., R. G. Smith, and H. J. Stocker, "Evidence for tunneling in reverse-biased III-V photodetector diodes," *A Phys. Lett.*, vol. 36, no. 7, pp. 580–582, 1980.

[14] P. A. Hiskett, J. M. Smith, G. S. Buller, and P. D. Townsend, "Low-noise single-photon detection at awavelength of 1.55 µm," *Electron. Lett.*, vol. 37, pp. 1081–1083, 2001.

[15] M. Kobayashi, S. Yamazaki, and T. Kaneda, "Planar InP/GaInAsP/GaInAs buried structure avalanche photodiode," *Appl. Phys. Lett.*, vol. 45, no. 7, pp. 759–761, 1984.

[16] Y. Kishi, K. Yasuda, S. Yamazaki, K. Nakajima, and I. Umebu, "Liquid-phase epitaxial growth of InP/GaInAsP/GaInAs buried-structure avalanche photodiodes," *Electron. Lett.*, vol. 20, no. 4, pp. 165–167, 1984.

[17] S. M. Sze and G. Gibbons, "Effect of junction curvature on breakdown voltage in semicondutors," *Solid-State Electron.*, vol. 9, no. 9, pp. 831–845, 1966.

[18] Z. I. Alferov, A. B. Zhuravlev, E. L. Portnoi, and N. M. Stel'makh, "Generation of picosecond pulses in hetero-lasers with a modulated durability," *Sov. Technol. Phys. Lett.*, vol. 12, pp. 452–458, 1996.

[19] J. M. Smith, P. A. Hiskett, I. Gontijo, L. Purves, and G. S. Buller, *Rev. Sci. Instrum*, vol. 72, pp. 2325–2329, 2001.

[20] S. Cova, M. Ghioni, A. Lacaita, C. Samori, and F. Zappa, "Avalanche photodiodes and quenching circuits for single photon detection," *Appl. Opt.*, vol. 35, pp. 1956–1976, 1996.

[21] M. A. Itzler, K. K. Loi, S. McCoy, N. Codd, and N. Komaba, "Manufacturable planar bulk-InP avalanche photodiodes for 10 Gb/s applications," in *Proc. 12th Ann. Meeting Lasers Electro-Optics Soc.*, 1999, Paper ThK0005, pp. 748–749.

[22] M. A. Itzler, K. K. Loi, S. McCoy, N. Codd, and N. Komaba, "Highperformance, manufacturable avalanche photodiodes for 10 gb/s optical receivers," in *Proc. Opt. Fiber Commun. Conf.*, Washington, DC, 2000, pp. 126–128.

[23] D. Stucki, G. Ribordy, A. Stefanov, H. Zbinden, J. G. Rarity, and T. Wall, "Photon counting for quantum key distribution with Peltier cooled InGaAs–InP APDs," *J. Mod. Opt.*, vol. 48, no. 13, pp. 1967–1981, 2001.

[24] S. R. Forrest and R. G. Smith, "Performance of In0.53Ga0.47As–InP avalanche photodiodes," *IEEE J. Quantum Electron.*, vol. QE-18, no. 12, pp. 2040–2048, Dec. 1982